\documentclass[a4paper,fleqn,usenatbib]{mnras}

\usepackage{newtxtext,newtxmath}

\usepackage[utf8]{inputenc}
\usepackage[T1]{fontenc}
\usepackage{ae,aecompl}

\usepackage{graphicx}	
\usepackage{amsmath}	
\usepackage{amssymb}	

\usepackage{graphicx}

\usepackage{bm}
\usepackage{tabularx}
\usepackage[roman]{parnotes}
\usepackage[normalem]{ulem}

\usepackage{siunitx}
\usepackage{color}

\newcommand{\vct}[1]{\bm{\mathrm{#1}}}
\newcommand{\mat}[1]{\mathsf{#1}}

\newcommand{\tnsf}{\mathcal{T}}

\newcommand{\fR}{\mathbb{R}}

\newcommand{\fromto}{\rightarrow}

\newcommand{\mean}[1]{\langle#1\rangle}

\DeclareMathOperator*{\argmin}{arg min}

\DeclareMathOperator{\arcosh}{arcosh}

\newcommand{\ud}{\mathrm{d}}

\newcommand{\chr}{\Gamma}

\newcommand{\derfrac}[2]{\frac{\ud #1}{\ud #2}}
\newcommand{\dnfrac}[3]{\frac{\ud^{#1} #2}{\ud {#3}^{#1}}}

\newcommand{\iprod}[2]{\left\langle#1,#2\right\rangle}
\newcommand{\norm}[1]{\left\|#1\right\|}
\newcommand{\abs}[1]{\left|#1\right|}

\newcommand{\eratio}{\xi}

\newcommand{\bigO}{O}

\newcommand{\arcmancer}{\mbox{\textsc{Arcmancer}}}
\newcommand{\harmpi}{\mbox{\textsc{harmpi}}}

\title[Semi-Riemannian barycentric interpolation]{Barycentric
interpolation on Riemannian and semi-Riemannian spaces}

\author[P.~Pihajoki, M.~Mannerkoski \& P.H.~Johansson]{
Pauli Pihajoki,$^{1}$\thanks{E-mail: pauli.pihajoki@iki.fi}
Matias Mannerkoski,$^{1}$
Peter H.~Johansson$^{1}$
\\
$^{1}$ University of Helsinki, Department of Physics, Gustaf Hällströmin
katu 2a, FI-00560 Helsinki, Finland
\\
}

\date{Accepted XXX. Received YYY; in original form ZZZ}

\pubyear{2018}

\begin{document}
\label{firstpage}
\pagerange{\pageref{firstpage}--\pageref{lastpage}}
\maketitle

\begin{abstract}
    Interpolation of data represented in curvilinear coordinates and
    possibly having some non-trivial, typically Riemannian or
    semi-Riemannian geometry is an ubiquitous task in all of physics.
    In this work we present a covariant generalization of the
    barycentric coordinates and the barycentric interpolation method for
    Riemannian and semi-Riemannian spaces of arbitrary dimension. We
    show that our new method preserves the linear accuracy property of
    barycentric interpolation in a coordinate-invariant sense.
    In addition, we show how the method can be used to
    interpolate constrained quantities so that the given constraint is
    automatically respected. We showcase the method with two
    astrophysics related examples situated in the curved Kerr spacetime.
    The first problem is interpolating a locally constant vector field,
    in which case curvature effects are expected to be maximally
    important.
    The second example is a General Relativistic Magnetohydrodynamics
    simulation of a turbulent accretion flow around a black hole,
    wherein high intrinsic variability is expected to be at least as
    important as curvature effects.
\end{abstract}




\section{Introduction}

Interpolation is necessary in a variety of physical problems,
both in modeling and the analysis of measurements. When the data are distributed
on a grid, the interpolation problem can be solved by the simplest of
methods such as $n$-dimensional linear interpolation. However, when the
data are
scattered, more general methods are required, such as kriging or
barycentric interpolation \citep{matheron1963, floater2006}.

In some cases the data can be both scattered and more
importantly distributed on a manifold with some intrinsic
geometry, such as for example the celestial sphere
\citep{kamionkowski1997,polarbear2014} or a planetary surface
\citep{colony1984,bindschadler1991,stohl1995}.
The data values may be also be constrained on some
submanifold with an induced geometry. A typical example is the velocity
field of matter, which in general or special relativity is
constrained to have unit norm everywhere. These complications can also
arise all at once, such as in simulations of strong gravity or of
cosmological scales. In both cases, the base
manifold is curved, and the velocity field is simultaneously constrained
\citep{etienne2012,adamek2014}. Finally, in addition to having intrinsic
geometry, the data might be known only as a distribution due to for
example measurement uncertainties. This final complication
requires statistical methods compatible with the data geometry
\citep[e.g.][]{pihajoki2017} and is beyond interpolation and the scope
of this work.

When the data, now assumed precisely known, are bound by some geometry,
the interpolation method used should respect this.
The intuitive motivation is two-fold: Firstly, the interpolated value
ought to be constructed using only the intrinsic variations within the
data, so that the result is independent of the choice of coordinates.
Secondly, the interpolated value should still be on the constraint
manifold in order to avoid having to project the value
back to the manifold.

A number of interpolation methods suited for scattered data on a
specific Riemannian manifold, the two-dimensional sphere $S^2$, have
been developed using a variety of approaches
\citep[e.g.][]{hardy1975,wahba1981,renka1984,lawson1984,pottmann1990,%
alfeld1996,cavoretto2010}.
Comparatively few algorithms have been invented for interpolation on
general Riemannian manifolds. However, a family of algorithms for
generalized Hermite(--Birkhoff) interpolation on closed, compact
Riemannian manifolds does exist
\citep{narcowich1995,dyn1999,allasia2018}, as well as a method for
nearest neighbour interpolation of vector and tensor fields
\mbox{\citep{sharp2019}}.
The authors know of no general purpose interpolation algorithms designed
specifically for scattered data on semi-Riemannian manifolds.

In this paper we present an intrinsic, coordinate-independent
generalization of the barycentric interpolation method to Riemannian and
semi-Riemannian spaces. The method is suitable for interpolation of
scattered tensorial data of any rank with or without constraints.
In Section~\ref{sc:bary}, we briefly review the concept of barycentric
coordinates and the standard barycentric interpolation method.
This is followed in Section~\ref{sc:generalization} by our
generalization. We show that the new method yields the linear precision
characteristic of barycentric interpolation in a coordinate-independent
manner, whereas the coordinate-only method fails to do so. We also
provide approximate formulae with which our method can be put into a
mathematically explicit form.
In Section~\ref{sc:numerical} we show numerical examples of the
behaviour of our algorithm in the case of a curved Kerr spacetime. We
provide here also the numerical implementation of the new method as a part of the
\arcmancer{}\footnote{\url{https://bitbucket.org/popiha/arcmancer}}
ray-tracing library \citep{pihajoki2018}.

\section{Barycentric coordinates and interpolation}\label{sc:bary}

\subsection{Barycentric coordinates}

Barycentric interpolation is based on the notion of barycentric coordinates.
Assume we have a convex polytope $P$ consisting of $N$ vertices
$V_1,\ldots,V_N$ in $n$ dimensions, with coordinates $\psi(V_i)=\vct{x}\in\fR^n$ in the
coordinate system $\psi$.
The corresponding barycentric coordinate functions
$\phi_1,\ldots,\phi_N:\fR^n\fromto\fR$
are then defined by the conditions $\phi_i\geq 0$ and
\begin{gather}
    \sum_{i=1}^{N} \phi_i(\vct{x}) \vct{x}_i = \vct{x}
    \label{eq:baryeq1} \\
    \sum_{i=1}^{N} \phi_i = 1, \label{eq:baryeq2}
\end{gather}
where $\psi(X)=\vct{x}\in\fR^n$ are the coordinates of the point $X$
under consideration.

If $N=n+1$, the barycentric coordinates can be directly determined from
equations \eqref{eq:baryeq1} and \eqref{eq:baryeq2}.
For $N>n+1$, the barycentric coordinate system is not unique.
A review of barycentric coordinate systems and
the methods to compute them is found in \cite{floater2015}.
Most of the coordinate systems were originally defined in 
two dimensions, but some, such as the maximum
entropy \citep{sukumar2004}, Wachspress
\citep{wachspress1975} and mean value coordinate systems \citep{floater2003}
generalize to higher dimensions.
Of these, the Wachspress and mean value coordinates use global geometric
properties such as areas, volumes or distances which make easy
generalization to semi-Riemannian spaces difficult. On the other hand,
the maximum entropy coordinates, briefly introduced in the following,
can be readily generalized to semi-Riemannian spaces.

\subsection{Maximum entropy coordinates}

The key idea of maximum entropy coordinates is to consider the
barycentric coordinates $\phi_i$ at a point $\vct{x}$ as a discrete
probability distribution. From this point of view, equation
\eqref{eq:baryeq1} is the statement $\mean{\vct{x}_i} = \vct{x}$, or
that the mean of $\vct{x}_i$ equals $\vct{x}$ over the distribution
$\{\phi_i\}$.
Another way to formulate this is to note that
equation~\eqref{eq:baryeq1} can be written as
\begin{equation}\label{eq:deltaxeq}
    \sum_{i=1}^N \phi_i(\vct{x}_i-\vct{x}) = \sum_{i=1}^N \phi_i
    \Delta\vct{x}_i = 0,
\end{equation}
where $\Delta\vct{x}_i = \vct{x}_i-\vct{x}$.
From this, we have correspondingly $\mean{\Delta \vct{x}_i} = 0$.
The coordinates $\phi_i$ can then be found by
maximizing the Shannon entropy $S(\{\phi_i\}) = -\sum \phi_i \ln \phi_i$,
subject to the constraints \eqref{eq:baryeq2} and \eqref{eq:deltaxeq}.
This can be achieved by introducing the Lagrange multipliers
$\alpha\in\fR$ and $\vct{\beta}\in\fR^n$, and maximizing the function
\begin{equation}\label{eq:consentropy}
    \begin{split}
        &S({\phi_i},\alpha,\vct{\beta}) = \\
        &\quad -\sum_{i=1}^N \phi_i\ln\phi_i
    + \alpha\left(\sum_{i=1}^N \phi_i - 1\right)
    + \vct{\beta}\cdot\sum_{i=1}^N \phi_i \Delta \vct{x}_i.
    \end{split}
\end{equation}
Setting the derivatives of equation \eqref{eq:consentropy} to zero
yields the equations
\begin{gather}
    Z = \sum_{i=1}^N Z_i = \exp(-\alpha+1) \label{eq:Zeq} \\
    Z_i = \exp(\vct{\beta}\cdot\Delta\vct{x}_i) \label{eq:Zieq} \\
    \phi_i = \frac{Z_i}{Z}. \label{eq:phiieq}
\end{gather}
The constraint \eqref{eq:deltaxeq} then gives
\begin{equation}
    \sum_{i=1}^N Z_i\Delta\vct{x}_i = 0,
\end{equation}
from which $\vct{\beta}$ and subsequently $\phi_i$ can be solved
directly. Alternatively, the solution can be obtained from the
equivalent minimization problem
\begin{equation} \label{eq:Zmineq}
    \vct{\beta} = \argmin_{\vct{\beta}'} \ln Z(\vct{\beta}'),
\end{equation}
which may be in some cases numerically easier \citep{sukumar2004,hormann2008}.

\subsection{Barycentric interpolation}

Barycentric coordinates have a natural application in interpolation.
Assume we wish to interpolate some field $f(\vct{x})$, where the
codomain of $f$ could be any vector space over $\fR$ in general, but for
typical physical applications would be $\fR$ itself or the space of
vectors or tensors at a point. We know the values of $f$ at the vertices
$\vct{x}_i$, and wish to obtain the interpolant $\widehat{f}(\vct{x})$
at $\vct{x}$.  When the barycentric coordinates $\phi_i$ of $\vct{x}$
have been obtained, the barycentric interpolant is computed from
\begin{equation}\label{eq:baryint}
    \widehat{f}(\vct{x}) = \sum_{i=1}^N \phi_i f(\vct{x}_i).
\end{equation}
The interpolation method \eqref{eq:baryint} turns out to have linear precision,
so that if $f(\vct{x})$ is linear in the coordinates
$\psi$, then $\widehat{f} = f$ inside the polytope $P$ \citep{floater2006}.

\section{Generalization to semi-Riemannian spaces}
\label{sc:generalization}

In the following, we now assume that the vertices $V_i$ are points on an
$n$-dimensional Riemannian or semi-Riemannian manifold $M$ with a metric
$g$ and some coordinate chart $\psi:M\fromto \fR^n$.
The data are taken to originate from an arbitrary rank $(r,s)$ tensor
field $f\in\tnsf^{r}_{s}(M)$, including scalar fields. Here
$\tnsf^r_s(M)$ is the space of all tensor fields on $M$, so that at each
vertex $V_i$ sits a tensor $f(V_i)\in T_{V_i}(M)^r_s$, where
$T_{V_i}(M)^r_s$ is the space of all rank $(r,s)$ tensors at point $V_i$.
We need a method to find the barycentric coordinates of $X$ with
respect to the $V_i$. We also need to transform the data from the
spaces $T_{V_i}(M)^r_s$ to the common space $T_{X}(M)^r_s$ to
compute the interpolant.

\subsection{Barycentric coordinates in curved spaces}

In order to find a generalization of barycentric coordinates to
semi-Riemannian spaces, we need to generalize the condition \eqref{eq:deltaxeq}
to curved spaces. This amounts to finding a reasonable generalization for the
difference of Cartesian position vectors $\Delta\vct{x}_i$.
An additional requirement for the object
sought after is that it should be defined locally at the point $X\in M$,
where we wish to compute the interpolant.

An object fulfilling these requirements is the tangent vector
$\vct{z}_i\in T_{X}(M)$ at $X$ of a geodesic
$\gamma_{\vct{z}_i}:\fR\fromto M$ for which $\gamma_{\vct{z}_i}(0) = X$
and $\gamma_{\vct{z}_i}(1) = V_i$.  Equivalently, the components of
$\vct{z}_i$ are the Riemann normal coordinates (RNC) of the point $V_i$
as developed around point $X$.  The vectors $\vct{z}_i$ are all defined
at $X$, and are a coordinate independent concept, as is the sum
$\sum_{i} \phi_i \vct{z}_i$.  Furthermore, for an Euclidean space with
Cartesian coordinates, we get precisely $\vct{z}_i = \Delta\vct{x}_i$.
While the RNC are often defined assuming an orthonormal basis of the
tangent space $T_XM$ at $X$ \citep[e.g.][]{oneill1983,lee1997}, this is
not strictly necessary and the components of $\vct{z}_i$ can be computed
in any local basis of $T_XM$ \citep{mtw1973}, such as the coordinate
vector basis of the chart $\psi$. In the following, we use the
coordinate vector basis of $\psi$ for all tensors.

Determining the vector $\vct{z}_i$ for a given $X$ and each $V_i$ in general
requires the solution of a boundary value problem
\begin{equation}
    \begin{split}
        &\frac{\ud^2 \gamma_{\vct{z}_i}(\lambda)^a}{\ud \lambda^2}
    + \chr^{a}_{bc} \frac{\ud \gamma_{\vct{z}_i}(\lambda)^b}{\ud \lambda}
    \frac{\ud \gamma_{\vct{z}_i}(\lambda)^c}{\ud \lambda} = 0 \\
        &\gamma_{\vct{z}_i}(0) = X \quad\quad \gamma_{\vct{z}_i}(1) = V_i.
    \end{split} \label{eq:bnd}
\end{equation}
This can be done numerically by combining a numerical ordinary differential
equation solver with a root-finding or optimization routine.

The new barycentric coordinate condition is now
\begin{equation}\label{eq:zeq}
    \sum_{i=1}^N \phi_i \vct{z}_i = 0,
\end{equation}
at the point $X$. The curved-space barycentric coordinates $\phi_i$ of $X$ can
again be found by maximizing the entropy
\begin{equation}
    \begin{split}
        &S(\{\phi_i\}, \alpha, \vct{\beta}) =\\
        &\quad
    -\sum_{i=1}^N \phi_i\ln\phi_i
    + \alpha\left(\sum_{i=1}^N \phi_i - 1\right)
    + \sum_{i=1}^N \langle\vct{\beta}, \phi_i\vct{z}_i\rangle,
    \end{split}
\end{equation}
where again $\alpha\in\fR$ but now $\vct{\beta}\in T^*_X(M)$ is an element
of the cotangent space at $X$, and the angle brackets denote the natural
pairing of tangent and cotangent spaces so that 
$\iprod{\vct{\beta}}{\vct{z}_i} = \beta_a z_i^a$
in the abstract index notation.
The solution to the maximization problem is identical to
equations \eqref{eq:Zeq}--\eqref{eq:Zmineq}, with $\Delta\vct{x}_i$
everywhere replaced with $\vct{z}_i$ and
$\vct{\beta}\cdot\Delta \vct{x}_i$ by $\langle
\vct{\beta},\vct{z}_i\rangle$.

It is interesting to note that to derive the barycentric coordinates for
a curved space, only the connection is required, for computing the
normal coordinates. A metric is not required, but for physically
interesting cases, a metric typically exists, and then the natural
choice for the connection is the metric (Levi--Civita) connection.

\subsection{Interpolation of unconstrained data}

For curved spaces, the interpolation formula \eqref{eq:baryint} cannot
be used directly. This is because the data $f(V_i)$ are defined in
different spaces with different base points $V_i$. To obtain the
interpolant $\widehat{f}$ at $X$, these data must first be transported
to $X$. For semi-Riemannian spaces, a natural solution is to parallel
transport the data from each $V_i$ back to $X$, using the same geodesic
$\gamma_{\vct{z}_i}$ used to define the RNC of each $V_i$.
This choice of curve is sufficient if we wish to retain the linear
accuracy property of barycentric interpolation, as is seen below in
Section~\ref{sc:proof}.

In the case of scalar data, after parallel transport we have just
equation \eqref{eq:baryint} again.
For
rank $(r,s)$ tensorial data, $f(V_i) = T(V_i)^{a_1\cdots a_r}_{b_1\cdots
b_s}$, we have to first solve the parallel transport problem
\begin{gather}
    \begin{split}
        &\frac{\ud T(\lambda)^{a_1\cdots a_r}_{b_1\cdots b_s}}{\ud \lambda} =
        \biggl(
         -\chr^{a_1}_{cd} T(\lambda)^{c\cdots a_r}_{b_1\cdots b_s}
        -\ldots
        -\chr^{a_r}_{cd} T(\lambda)^{a_1\cdots c}_{b_1\cdots b_s}
        \\
        &\quad +\chr^{c}_{b_1d} T(\lambda)^{a_1\cdots a_r}_{c\cdots b_s} 
        +\ldots
        +\chr^{c}_{b_sd} T(\lambda)^{a_1\cdots a_r}_{b_1\cdots c}
        \biggr) v_i(\lambda)^d
    \end{split} \label{eq:partrans1} \\
    T(1)^{a_1\cdots a_r}_{b_1\cdots b_s} = T(V_i)^{a_1\cdots
    a_r}_{b_1\cdots b_s}, \label{eq:partrans2}
\end{gather}
where $v_i(\lambda)^a = \ud \gamma_{\vct{z}_i}(\lambda)^a/\ud\lambda$,
to obtain the parallel transported 
$\widetilde{T}^{a_1\cdots a_r}_{b_1\cdots b_s} =
T(0)^{a_1\cdots a_r}_{b_1\cdots b_s}$.
After the parallel transport, all quantities are now defined in the same
space at the point $X$,
and we can use equation \eqref{eq:baryint} to obtain
\begin{equation}\label{eq:unconsint}
    \widehat{T} = \sum_{i=1}^N \phi_i \widetilde{T}_i,
\end{equation}
where the tensor indices have been suppressed.

\subsection{Interpolation of constrained data}

Often the field $f$ to be interpolated has additional constraints. The
archetypal example in relativity is the four-velocity field $\vct{u}$ of
matter, which has to fulfill $g_{ab}u^a u^b = 1$ everywhere. The
existence of a constraint restricts the field $f(p)$ to a
subspace $S_p\subset T_p(M)^r_s$ of the space of all rank $(r,s)$
tensors at each point $p\in M$.
Here we assume that $S_p$ is a (semi-)Riemannian
submanifold of $T_p(M)^r_s$ for each $p$, with a metric $g_S$,
possibly induced from the metric $g$.
If the constraint is compatible with parallel transport, we can still
obtain the parallel transported data $\widetilde{T}_i$ at the
interpolation point $X$.
However, now the direct
interpolant $\widehat{T}$ given by equation~\eqref{eq:unconsint} is in
general not a member of the constraint space $S_p$. For example in the
case of unit vector fields, the magnitude of the interpolant is always
less than or equal to unity in Riemannian spaces, and greater than or
equal to unity for timelike unit vector fields in Lorentzian spaces.
This is because the barycentric interpolant is a convex combination.

The problem can be solved by using the method to compute the barycentric
coordinates in reverse.
The constraint submanifold $S_X\subset T_X(M)^r_s$ has a geometry
defined by the metric $g_S$ by assumption. We can find the interpolant
$\widehat{T}$ by requiring that
\begin{equation}\label{eq:consint}
    \sum_{i=1}^N \phi_i\widetilde{t}_i = 0,
\end{equation}
where $\widetilde{t}_i$ are now the Riemann normal coordinates of the
data $\widetilde{T}_i$ as developed at point $\widehat{T}$. In the case
of no constraints, or $S_p = T_p(M)^r_s$, the geometry is flat since
$T_p(M)^r_s$ is a vector space isomorphic to $\fR^{r\times s}$, and we have
$\widetilde{t}_i=\widehat{T}_i-\widehat{T}$.
In this case, the equation \eqref{eq:consint}
reduces to equation \eqref{eq:unconsint}.

\subsection{Proof of linear accuracy}
\label{sc:proof}

In the following, we show that the interpolation method
presented in this paper is the correct covariant generalization of
barycentric interpolation in the sense that it preserves the property of
linear accuracy in curved spaces in a coordinate-independent sense,
whereas the usual coordinate space method fails to do so.
Here we assume that the vertices $V_i$ where the
data are located at are contained in a relatively small region compared
to the curvature scale, as well as the scale of variation of the field
from which the data are sampled.
The accuracy of the curved space barycentric
interpolation scheme described above can then be compared to the
standard coordinate-only barycentric interpolation by using Taylor
expansions up to second order.

Let the coordinates of the interpolation point $X$ be $\vct{x}$ and the
coordinates of the vertices $V_i$ be $\vct{x}_i$, and let us write
$\Delta\vct{x}_i = \vct{x}_i - \vct{x}$. By expanding the geodesic
$\gamma_i$
connecting $X$ and $V_i$ in a Taylor series, we find the following
relations between the normal coordinates $\vct{z}_i$ of the vertex $V_i$
and the coordinate differences $\Delta\vct{x}_i$
(see also \cite{brewin2009})
\begin{align}
    z_i^a &= \Delta x_i^a + \frac{1}{2}\chr^a_{bc}\Delta x_i^b \Delta x_i^c  
    + \bigO(\Delta \vct{x}_i^3) \label{eq:dx2rnc_ex} \\
    \Delta x_i^a &= z_i^a - \frac{1}{2}\chr^a_{bc}z_i^b z_i^c
    + \bigO(\vct{z}_i^3), \label{eq:rnc2dx_ex}
\end{align}
where the $\bigO$-notation indicates that the quantities are given to
second order in the components of $\Delta\vct{x}_i$ and $\vct{z}_i$,
which are by assumption small.
Here and throughout the rest of this section the Christoffel symbols and
their derivatives are computed in the original coordinate basis $\psi$.

Let us then assume that the data to be interpolated is represented by a
vector field $\vct{u}\in \tnsf^1_0(M)$. The following derivation works
for general tensor fields as well, but the algebraic complexity grows
significantly with each additional index.
We evaluate $\vct{u}$ at $V_i$ by propagating it from $X$
along the geodesic $\gamma_i$. The derivatives of $\vct{u}$ with respect
to the curve parameter $\lambda$ along $\gamma_i$ are then
\begin{align}
    \derfrac{u^a}{\lambda}(\lambda) &=  \Bigl(u^a_{;c} - \chr^a_{bc}
    u^b\Bigr) \derfrac{\gamma_i^c}{\lambda} \\
    \begin{split}
    \dnfrac{2}{u^a}{\lambda}(\lambda) &= \Bigl(
    u^a_{;cd} - \chr^a_{bc} u^b_{;d} - \chr^a_{bd} u^b_{;c} -
        \chr^a_{bc,d}u^b \\
    &\quad + \chr^a_{ce}\chr^e_{bd} u^b + \chr^a_{be}\chr^e_{cd}u^b
        \Bigr) \derfrac{\gamma_i^c}{\lambda}\derfrac{\gamma_i^d}{\lambda},
    \end{split}
\end{align}
where we have used the notation $u^a_{;cd} = \nabla_d\nabla_c u^a$ for
covariant derivatives, which indicate the `true' change in
$\vct{u}$ as opposed to purely coordinate or curvature related effects.
Since $V_i = \gamma_i(1)$, we can express $\vct{u}$ at $V_i$ to second order in
$\vct{z}_i$ through
\begin{equation}
    \begin{split}
        u_i^a &= u_0^a + \derfrac{u^a}{\lambda}(0) 
        + \frac{1}{2}\dnfrac{2}{u^a}{\lambda}(0) \\
        &= u_0^a + (u^a_{;c} - \chr^a_{bc}u_0^b)z_i^c \\
         &\quad\quad +\frac{1}{2}\Bigl(
        u^a_{;cd} - 2\chr^a_{bc} u^b_{;d} - \chr^a_{bc,d}u_0^b \\
        &\quad\quad\quad
        +\chr^a_{ce}\chr^e_{bd} u_0^b + \chr^a_{be}\chr^e_{cd}u_0^b
        \Bigr) z_i^c z_i^d
    \end{split}
\end{equation}
where we have used the symmetry of $z_i^c z_i^d$, and where now all
Christoffel symbols and derivatives are evaluated at $X$.

When the data $\vct{u}_i$ are parallel transported back to $X$ we get
\begin{equation}
    \widetilde{u}_i^a =
        u_0^a + z_i^c u^a_{;c} +\frac{1}{2} z_i^c z_i^d u^a_{;cd}
        + \bigO(z_i^3).
\end{equation}
The curved space barycentric interpolant, equation~\eqref{eq:unconsint},
is then, to second order in $\vct{z}_i$, given by
\begin{equation}
    \label{eq:curved-2nd}
    \widehat{u}^a = \sum_{i=1}^N \phi_i \widetilde{u}_i^a
    = u_0^a + \frac{1}{2} u^a_{;cd} \sum_{i=1}^N \phi_i z_i^c z_i^d,
\end{equation}
whereas the usual geometry-ignorant (`flat') barycentric interpolation yields
similarly to second order
\begin{equation}
    \label{eq:flat-2nd}
    \begin{split}
        \widehat{u}^a &= \sum_{i=1}^N \chi_i u_i^a\\
        &= u_0^a + \frac{1}{2}\left(u^a_{;cd} + C^a_{cd}(u) \right)
            \sum_{i=1}^N \chi_i \Delta x_i^c \Delta x_i^d,
    \end{split}
\end{equation}
where
\begin{equation}
    C^a_{cd}(u) = \chr^b_{cd}u^a_{;b} - 2\chr^a_{bc} u^b_{;d} 
 + \chr^a_{ce}\chr^e_{bd}u_0^b -\chr^a_{bc,d} u_0^b
\end{equation}
is the contribution from curvature, and
we have used $\chi_i$ for the `flat' barycentric coordinates.
Here the curved space barycentric coordinates $\phi_i$ fulfill
equation~\eqref{eq:zeq}, whereas the `flat' barycentric coordinates
$\chi_i$ solve equation~\eqref{eq:deltaxeq} instead.

Comparing equations \eqref{eq:curved-2nd} and \eqref{eq:flat-2nd}, we
can immediately point out our key results.
First, when there is no curvature in the space or the
coordinate system, so that $\chr^a_{bc} = \chr^a_{bc,d} = 0$, both
methods give equivalent results and are exact for locally linear fields,
as expected.
Furthermore, we see that if the vector field $\vct{u}$ is locally linear, so that
$u^a_{;cd}=0$, then the curved space method, equation
\eqref{eq:curved-2nd} gives exact results, whereas the flat method,
equation \eqref{eq:flat-2nd}, does not. In fact, the flat method fails
to give correct results even for locally \emph{constant} fields, i.e.
$u^a_{;cd}=u^a_{;c}=0$, due to the introduction of curvature effects. It should be
emphasized here that the result in equation \eqref{eq:curved-2nd}
requires both aspects of the new interpolation algorithm: computing the
barycentric coordinates using the Riemann normal coordinates
\emph{and} the parallel transport of the data back to the interpolation
point. As such, we conclude that the method presented here is a
correct covariant generalization of the flat space barycentric
interpolation method.

\subsection{Approximate formulae}\label{sc:approximate}

For generic problems, the interpolation method described in
Section~\ref{sc:generalization} cannot be put into an explicit form.
Instead, several steps of numerical computation are required. Firstly,
finding the Riemann normal coordinates of the vertices requires several
solutions of a boundary value problem \eqref{eq:bnd}
for the geodesic equation.
After this, the maximum entropy procedure to obtain the barycentric
coordinates requires a solution to an optimization problem, equation
\eqref{eq:Zmineq}, or a root finding
problem, equation \eqref{eq:zeq}. Finally, if the data are constrained, a
combined solution of finding Riemann normal coordinates and root
finding, equation \eqref{eq:consint}, is required. This can amount to a
large computational cost per single interpolation. However, if the size
of the region containing the vertices $V_i$ is comparatively small, and
the curvature of the space is likewise small, explicit forms for the
some sub-steps of the interpolation method can be derived in an approximate
form.

The problem of finding Riemann normal coordinates can be transformed
into an explicit form using series approximations \citep{brewin2009}.
For example, to second order, we have the formulae \eqref{eq:dx2rnc_ex}
and \eqref{eq:rnc2dx_ex}.

The minimization problem \eqref{eq:Zmineq} can also be explicitly solved
in the special case where the interpolation point $X$ is near the
barycentre of the vertices $V_i$. In this case we have $\vct{\beta}\sim 0$,
and the barycentric coordinate condition \eqref{eq:zeq} gives
\begin{equation}
    \sum_{i=1}^N (1 + \langle\vct{\beta},\vct{z}_i\rangle) \vct{z}_i +
    \bigO(\vct{\beta}^2) = 0.
\end{equation}
Discarding the higher order terms, this equation can be directly solved
through
\begin{gather}
    \vct{\beta} = -(\mat{Z}^T\mat{Z})^{-1} \mat{Z}^T \vct{1}, \\
    \intertext{where}
    \mat{Z}^T = \begin{pmatrix} \vct{z}_1 & \cdots & \vct{z}_N \end{pmatrix} 
    \quad \text{($n\times N$ real matrix)} \\
    \vct{1}^T = (1,\ldots,1)
    \quad \text{($1\times N$ row vector)}.
\end{gather}
This can be seen to be equivalent to the least squares solution of the equation
$\mat{Z}\vct{\beta} = -\vct{1}$.

The parallel transport problem, equations
\eqref{eq:partrans1}--\eqref{eq:partrans2} can also be solved
approximately using series methods. If the RNC of a vertex $V$ are
$\vct{z}$, the data at $V$ is a rank $(r,s)$ tensor 
$T^{a_1\cdots a_r}_{b_1\cdots b_s}$
and $\chr^a_{bc}$ are the Christoffel symbols evaluated at the
interpolation point $X$, then to first order the parallel transported
tensor $\widetilde{T}$ at $X$ is
\begin{equation}
    \begin{split}
        \widetilde{T}^{a_1\cdots a_r}_{b_1\cdots b_s} &= 
        T^{a_1\cdots a_r}_{b_1\cdots b_s}
         +\chr^{a_1}_{cd} T^{c\cdots a_r}_{b_1\cdots b_s} z^d
        +\ldots
        +\chr^{a_r}_{cd} T^{a_1\cdots c}_{b_1\cdots b_s} z^d
        \\
        &\quad -\chr^{c}_{b_1d} T^{a_1\cdots a_r}_{c\cdots b_s} z^d
        -\ldots
        -\chr^{c}_{b_sd} T^{a_1\cdots a_r}_{b_1\cdots c} z^d.
    \end{split}
\end{equation}
And in particular for ubiquitous vectorial data, we have
\begin{equation}
    \widetilde{T}^a = T^a + \chr^a_{bc} T^b z^c.
\end{equation}

For interpolating constrained data, there is naturally no generic
explicit solution. This is also true for such physically motivated
simple cases as timelike unit vector fields on Lorentzian manifolds, but for
these at least the solution can be condensed to a single implicit
equation.
Assume that we have previously derived the barycentric coordinates
$\phi_i$ of the interpolation point $X$, and we have several timelike unit
vectors $\vct{v}_i$ that have been parallel transported from the
vertices to $X$.
We wish to find the interpolant $\vct{v}$ fulfilling equation
\eqref{eq:consint}. We further assume that the metric at $X$ is in
the Minkowski form, which can always be achieved by e.g.\
orthonormalizing the coordinate frame.
In the space of timelike unit vectors at $X$, the Riemann normal
coordinates of a vector $\vct{v}_i$ with respect to the vector $\vct{v}$
can now be found in the following manner.
First, set
\begin{align}
    \vct{u} &= \vct{v}_i-\vct{v} - \iprod{\vct{v}_i-\vct{v}}{\vct{v}}\\
    \vct{\widehat{u}} &=
    \frac{\vct{u}}{\sqrt{-\iprod{\vct{u}}{\vct{u}}}},
\end{align}
so that $\vct{\widehat{u}}$ is the normalized part of
$\vct{v}_i-\vct{v}$ orthogonal to $\vct{v}$.
Now
\begin{equation}
    \vct{\gamma}(\omega) = \cosh(\omega) \vct{v} + \sinh(\omega)
    \vct{\widehat{u}}
\end{equation}
is a geodesic in the velocity space, with $\vct{\gamma}(0) = \vct{v}$
and $\vct{\gamma}(\omega^*) = \vct{v}_i$, from which we get $\omega^* =
\arcosh(\iprod{\vct{v}}{\vct{v}_i})$. Thus the Riemann normal
coordinates $\vct{t}_i$ of $\vct{v}_i$ developed at point $\vct{v}$ are
\begin{equation}
    \vct{t}_i = \arcosh(\iprod{\vct{v}}{\vct{v}_i}) \vct{\widehat{u}}. 
\end{equation}
The desired unit length interpolant $\vct{v}$ is then found by solving
$\sum_i \phi_i \vct{t}_i = 0$, a system of $n$ non-linear equations for the
$n$ components of $\vct{v}$.

\subsection{When are curvature effects important?}\label{sc:curvature}

Since the curved space computation is potentially more numerically
demanding, it would be useful to know when exactly we can expect to
benefit from such a procedure.
The equation for the second order error of the flat barycentric
interpolation, equation \eqref{eq:flat-2nd}, provides an estimate for when
curvature effects should be taken into account.

For scalar fields, we see that the only difference between flat and curved
space barycentric interpolation is in the computation of the barycentric
coordinates. This in turn depends on the difference between $\Delta
\vct{x}_i$, the coordinate differences between the vertex and the
interpolation point, and $\vct{z}_i$, the Riemann normal coordinates of
the vertex with respect to the interpolation point. From equations
\eqref{eq:dx2rnc_ex} and \eqref{eq:rnc2dx_ex} we see that
this difference is proportional to the components of the connection and
the amount of coordinate difference. Thus we can say that when the
values of $\frac{1}{2}\chr^a_{bc}\Delta x_i^b \Delta x_i^c$ are small
compared to $\Delta x_i^a$ itself, the curvature effects can be safely
ignored for all scalar field interpolation.

For vector and tensor fields, the situation is more complicated.
Firstly, we have the condition obtained above for scalar fields.  In
addition, from equation \eqref{eq:flat-2nd} we can deduce that curvature
effects are likely to be important when the natural variability,
represented by the second order covariant derivatives of the field, is
of the same order as the curvature error contribution $C^a_{cd}(u)$.

In the following section, we will numerically investigate situations
in curved spaces where the curvature effects are either crucial or of
limited importance.

\section{Numerical examples}\label{sc:numerical}

\subsection{Locally constant vector fields}\label{sc:locally_const}

To clearly illustrate the difference between the new method and coordinate space
methods, we investigated numerically the interpolation error in a curved
spacetime as a function of the size of the interpolation region.
We computed the interpolation
error in the squared norm and the components of a locally constant
unit-norm vector field $\vct{u}$ in the Kerr spacetime \citep{kerr1963},
with a mass parameter $M=1$ and a dimensionless spin parameter $\chi=0.99$.
Since the vector field was taken to be locally constant, we have
$u^a_{;bc} = 0$, and by the discussion in Section~\ref{sc:curvature}, the
error caused by neglecting curvature should be significant.

The vector field was computed in the outgoing Cartesian Kerr--Schild
coordinates $(t,x,y,z)$ \citep{kerr1965}. Local constantness was
achieved up to numerical precision by parallel transporting a vector
$\vct{u}_0 = (g_{tt}^{-1/2}, 0, 0, 0)$ from the interpolation point to
the data vertices using the \arcmancer{} code \citep{pihajoki2018}.
The vertices were defined
to span the hypercube $[0,L]\times[10,10+L]\times[0,L]^2$, where $L$
determines the size of the interpolation region. The interpolation point
was set in the coordinate center of the hypercube.

\begin{figure}
    \begin{center}
    \includegraphics[width=\columnwidth,keepaspectratio]{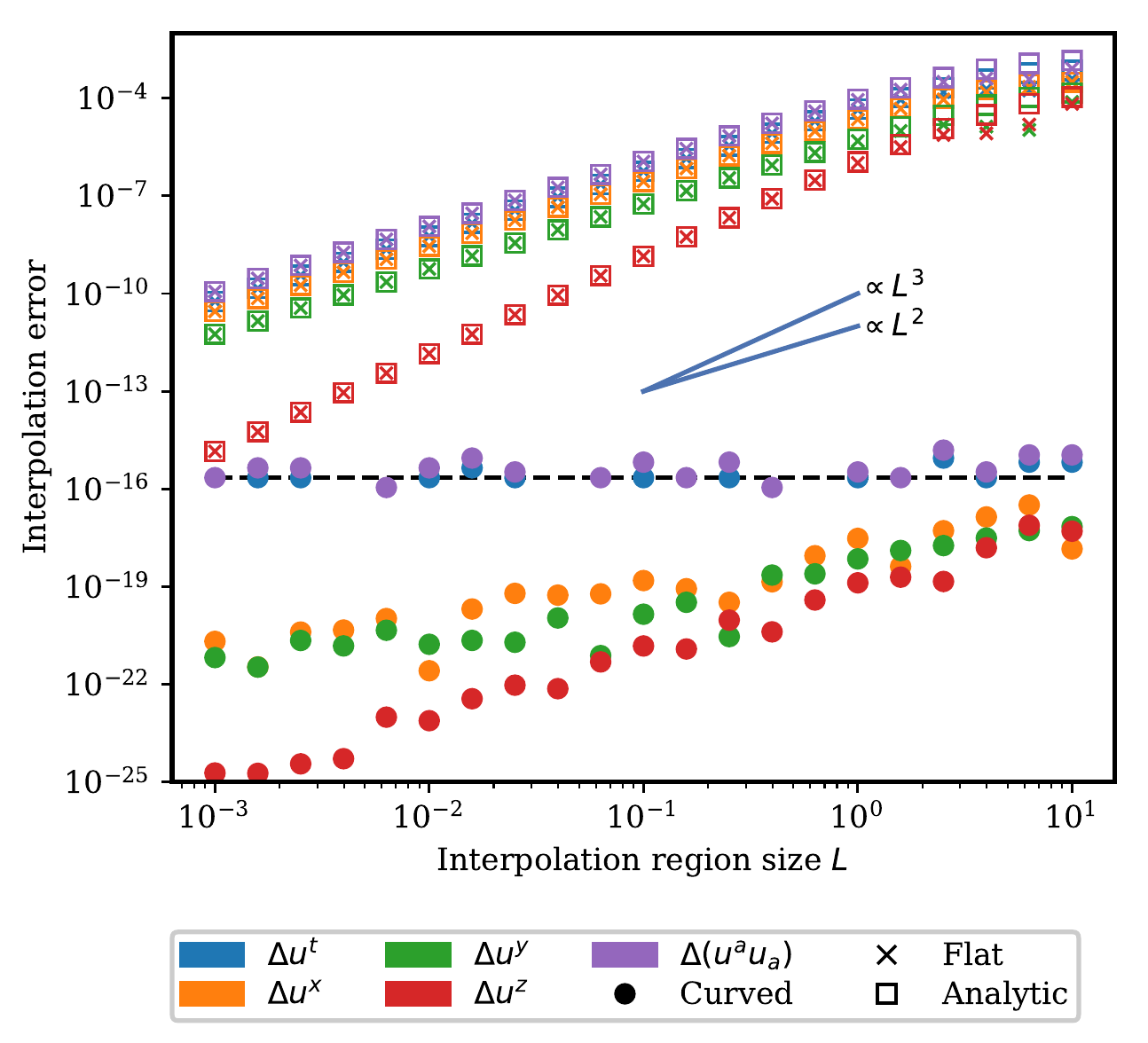}
    \caption{Interpolation error for a locally constant vector field as
    a function of the interpolation grid cell size $L$.
    The curved space barycentric method is indicated by
    filled circles, and the flat space barycentric method by crosses.
    Open squares indicate the analytic error estimate from
    equation~\eqref{eq:flat-2nd}. The black dashed line indicates the
    double precision floating point machine epsilon level.
    The figure includes two lines
    proportional to $L^2$ and $L^3$, drawn to guide the eye.
    }\label{fig:error}
    \end{center}
\end{figure}

Figure~\ref{fig:error} presents the absolute interpolation error of the
curved space barycentric interpolation method, labelled `Curved',
and the coordinate space barycentric interpolation method, labelled
`Flat'. We also computed the results of a standard $n$-linear coordinate
interpolation method, which were identical to the flat space barycentric
result to within numerical precision, as is expected when the data
points lie on a regular grid \citep{sukumar2004}.
In addition, we computed the analytic error estimate for the flat
barycentric method given by equation~\eqref{eq:flat-2nd}. The analytic
estimate was found to be in excellent agreement with the numerical
results up to interpolation region size of $\sim 1$, where higher order
corrections can be expected to become significant. From the figure we
see that the interpolation method presented here has essentially no
error, with only round-off error from the numerical geodesic integration
and optimization present.
The level of numerical error seen in Figure~\ref{fig:error} 
can be understood by considering
the following rough estimate. If an exact process starting from some
$x_0+\Delta$ and progressing to $x_0$ in $N$ small steps is approximated
by an iterative numerical operation proceeding by consecutive
differences, the resulting error is at most $\sim 2N\epsilon x_0 +
N\epsilon\Delta + \bigO(\epsilon^2)$, where $\epsilon$ is the machine
epsilon, or the largest possible relative error when rounding to one
\citep{goldberg1991}. In this case, $\Delta$ is small and for the $u^t$
component $x_0 \approx 1$, and for the $u^x$, $u^y$ and $u^z$ components 
$x_0 = 0$, which yields an estimate consistent with the errors seen in
Figure~\ref{fig:error}, since we can expect $N$ to increase with $L$.
In contrast to the curved space method, the standard
coordinate-space methods are strictly bound by a non-zero error that
scales at least quadratically with the size of the interpolation region.
This result numerically illustrates the content of equations
\eqref{eq:curved-2nd} and \eqref{eq:flat-2nd}.

\subsection{Interpolating a GRMHD simulation}

We also investigated a situation, where the natural variability of the
interpolated data can be expected to be high, and to possibly even surpass
the effects of non-zero curvature. For this purpose, we chose a dataset
consisting of a General Relativistic Magnetohydrodynamics (GRMHD)
simulation of a turbulent, magnetized accretion flow around a Kerr black
hole.

We used the \harmpi{}\footnote{%
\harmpi{} is freely available at
\url{https://github.com/atchekho/harmpi}.%
}
code \citep{gammie2003,noble2006} to run a three-dimensional simulation
of a magnetized plasma torus around a Kerr black hole, with
dimensionless spin of $\chi=0.5$. The initial conditions used were the
standard initial conditions provided by the \harmpi{} code, describing
the Fishbone--Moncrief solution \citep{fishbone1976} of a poloidally
magnetized plasma torus with a pressure maximum at a radius $r=12M$,
where $M$ is the black hole mass. The number of (equidistant) grid
points in the internal $(X_1,X_2,X_3)$ coordinates was set to
$(126,96,96)$. These coordinates correspond in a non-linear fashion to
the spherical Kerr-Schild radial, polar and azimuthal coordinates
$(r,\theta,\phi)$.

We evolved the simulation until $t=4250\,M$. From the final snapshot
we created an output, downsampled by a factor of two by discarding data at
every other grid point. Using the downsampled
output, we interpolated the values of plasma density, internal energy,
velocity and magnetic field at such points where the data had been
discarded during downsampling, using both the method described in this paper and standard
$n$-linear interpolation in the coordinate space. In the $n$-linear
case, the velocity vectors were normalized to unit length after
interpolation to obtain parity with the constrained barycentric method.
All interpolation was done using the Cartesian ingoing Kerr-Schild
coordinates \citep[see e.g.][]{carter1968}, in order not to introduce
additional curvature from the use of a spherical coordinate chart.

\begin{figure*}
    \begin{center}
        \mbox{
            \includegraphics[width=0.5\textwidth]{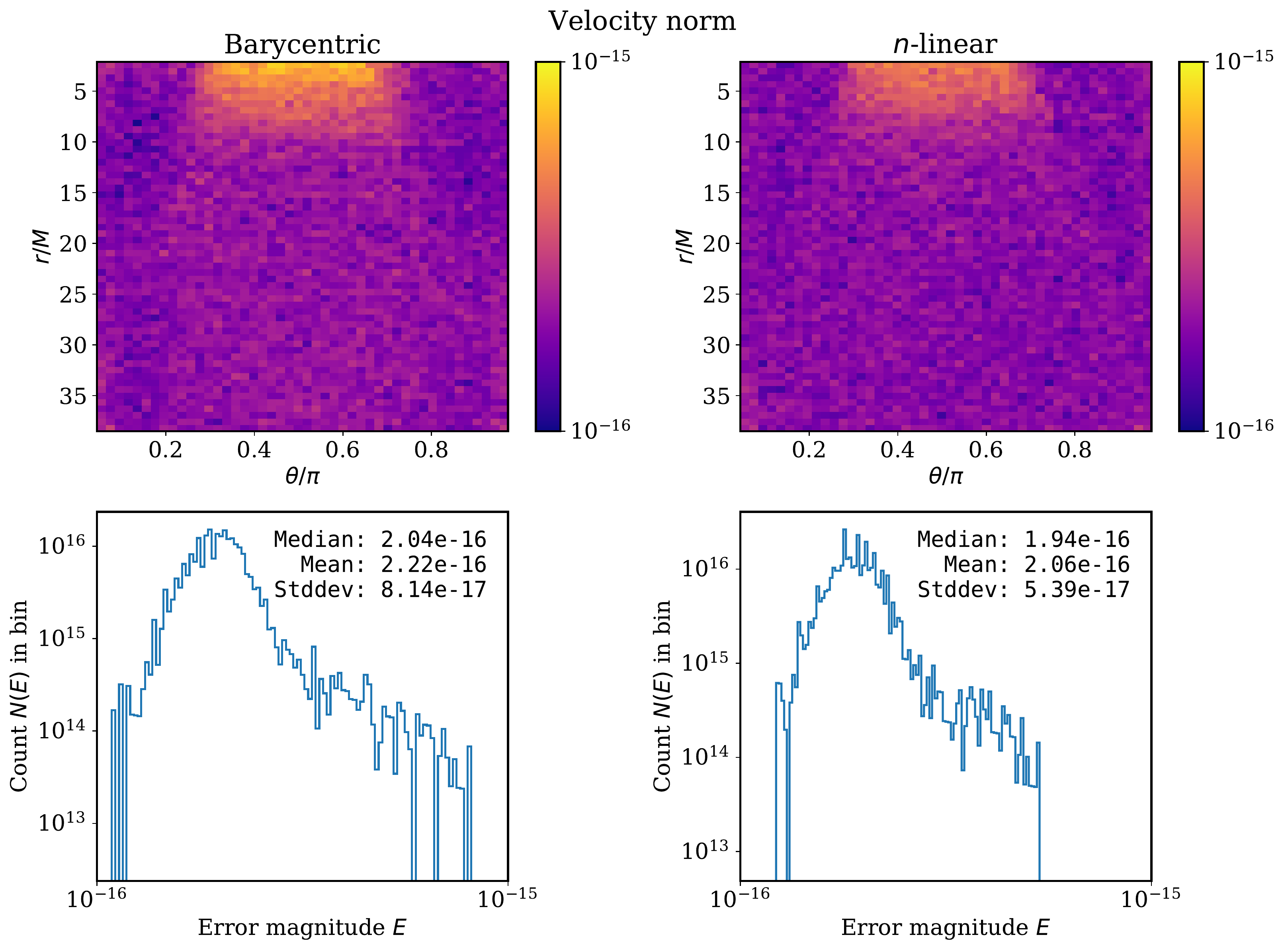}
            \includegraphics[width=0.5\textwidth]{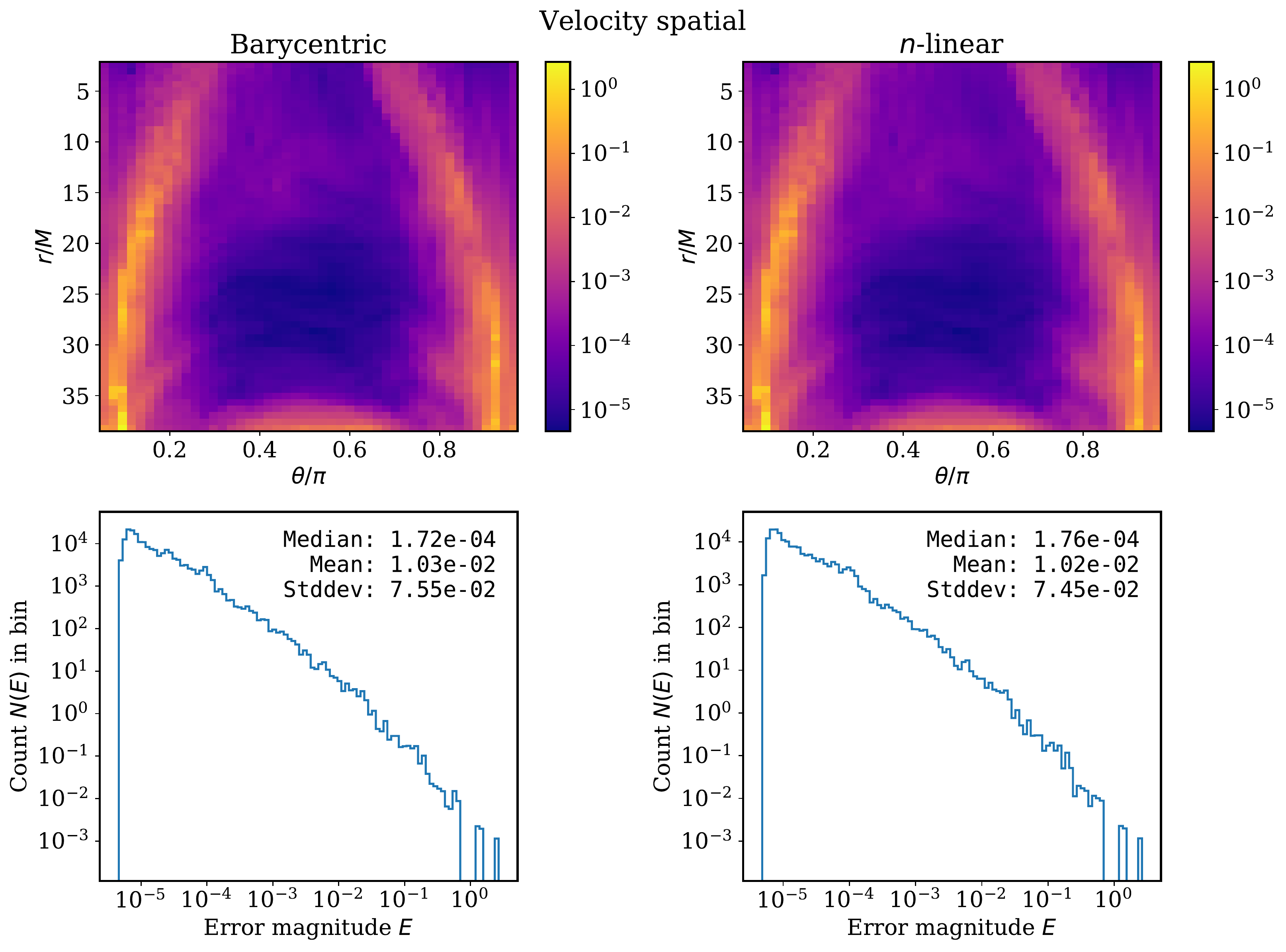}}
        \mbox{
            \includegraphics[width=0.5\textwidth]{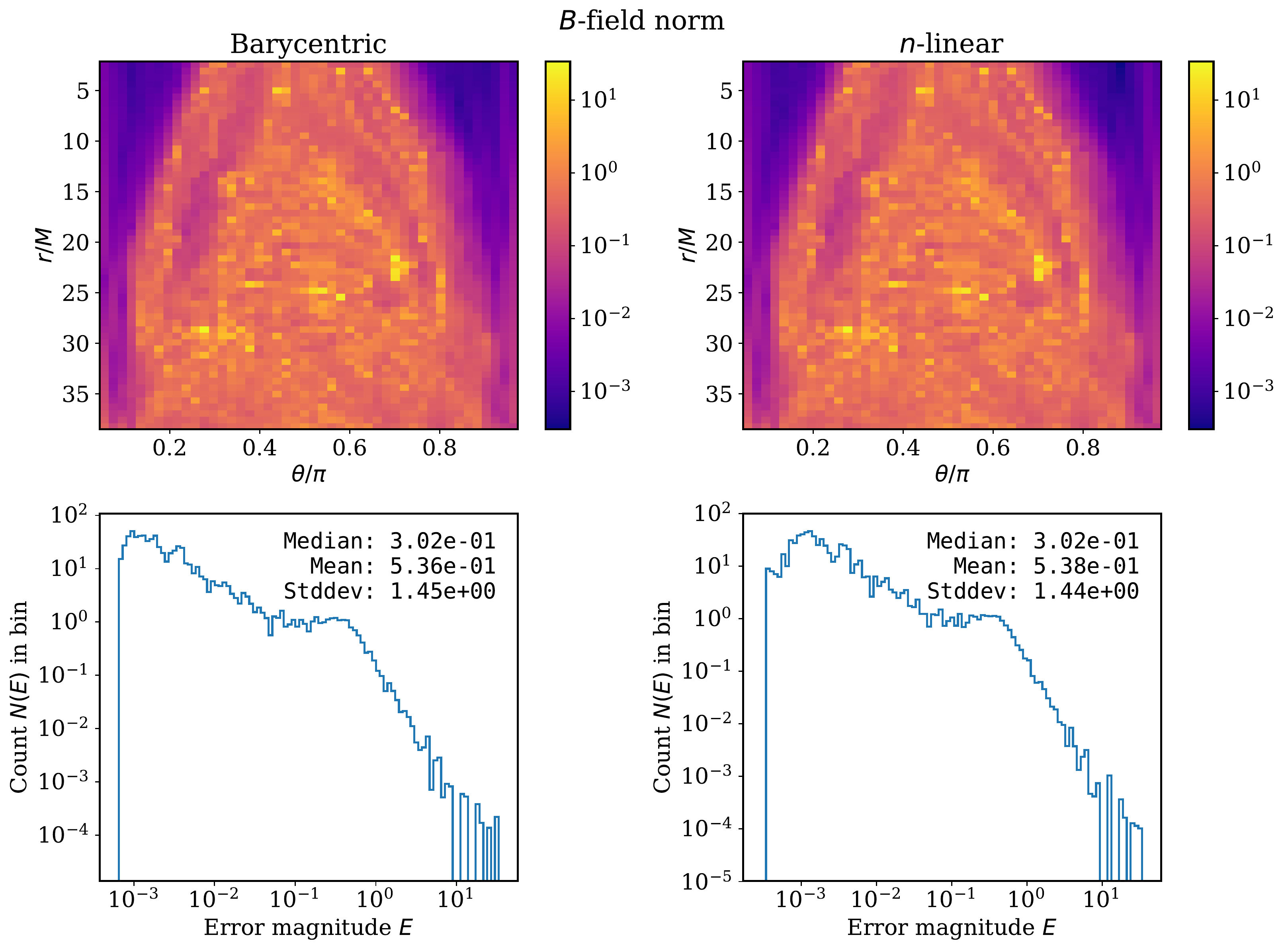}
            \includegraphics[width=0.5\textwidth]{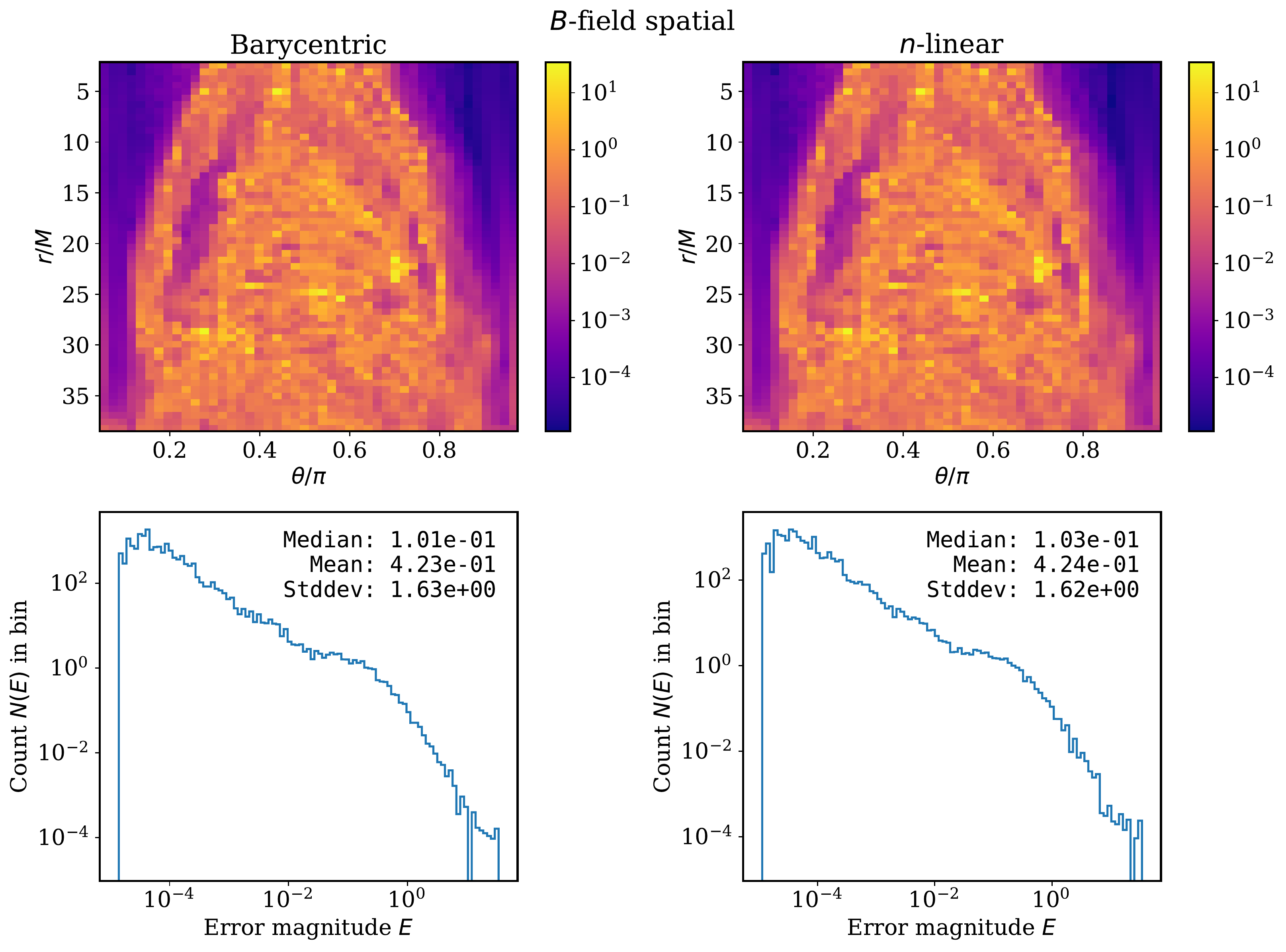}}
        \mbox{
            \includegraphics[width=0.5\textwidth]{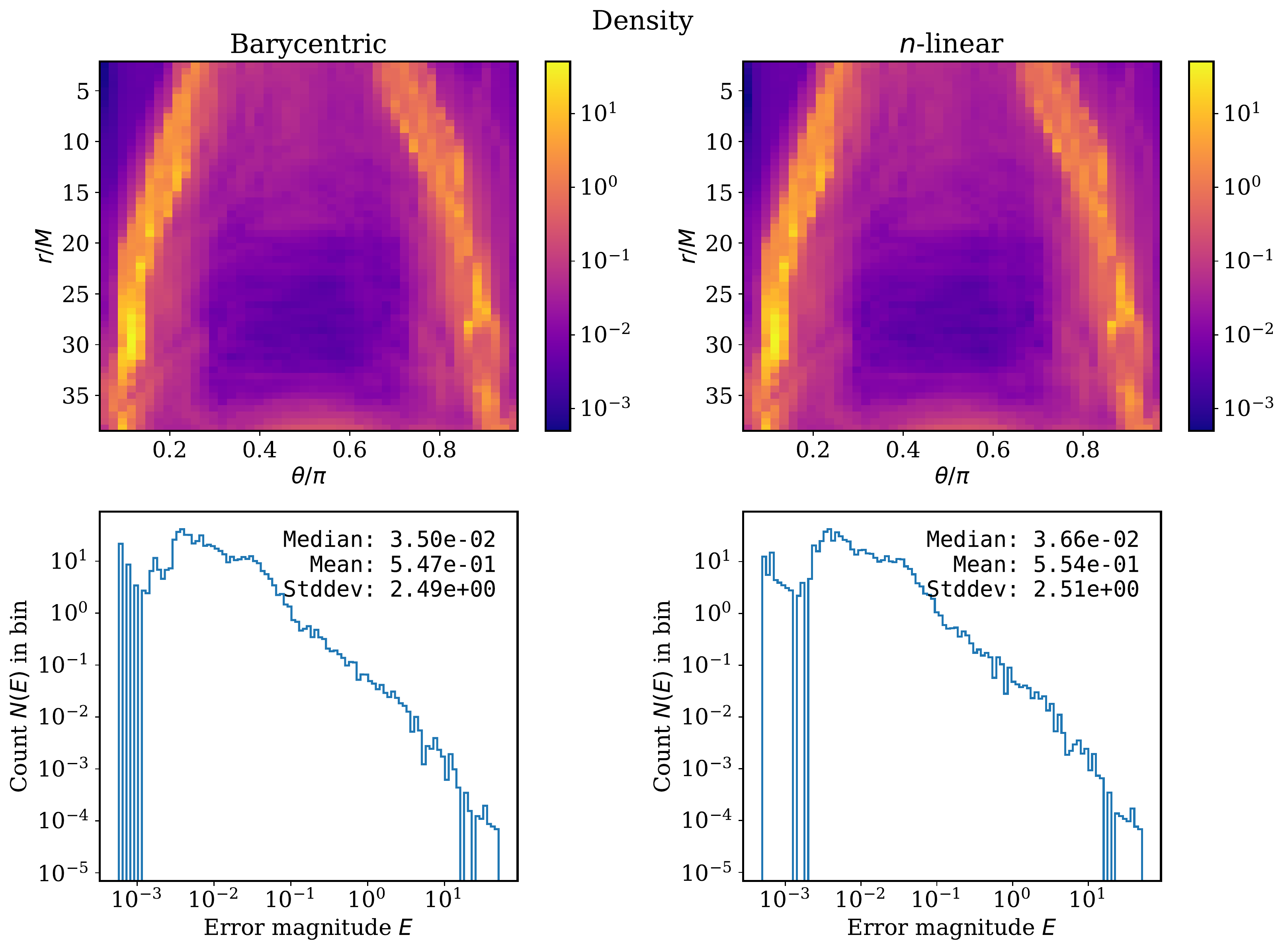}
            \includegraphics[width=0.5\textwidth]{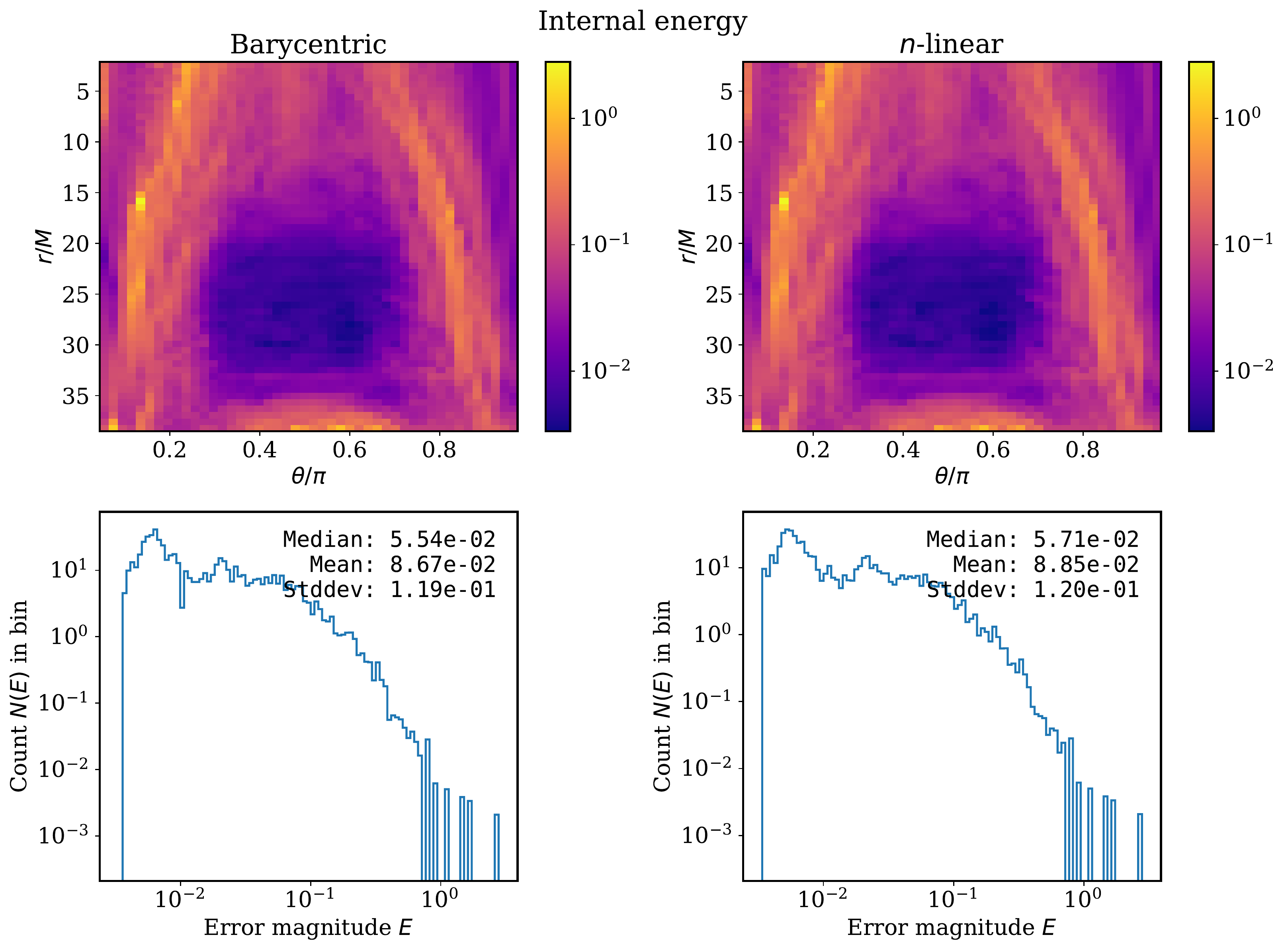}}
        \caption{Absolute relative errors in the squared norms and
        spatial components of the velocity and magnetic field
        ($B$-field), as well as density and internal energy, from left
        to right and top to bottom. All errors have been averaged over the
        azimuthal angle $\phi$.
        Each quantity has a $4\times4$ panel
        where the top panels show a heatmap of the distribution of the
        error in the $(r,\theta)$ plane and bottom panels display
        histograms of the distribution of the relative error magnitudes.
        The leftmost two panels give the results corresponding to the
        barycentric method presented in this paper, and the rightmost
        two panels give the results of a standard $n$-linear coordinate
        interpolation method, combined with a normalization of the
        resulting velocity vector.}
    \label{fig:grmhd_plots}
    \end{center}
\end{figure*}

The interpolated
values were then compared with the known values of the original
snapshot, and the differences averaged over the azimuthal angle $\phi$,
corresponding to averaging around the spin axis of the black hole. The
results are shown in Figure~\ref{fig:grmhd_plots}. In the Figure, we plot the
relative errors in the squared norm and spatial components of the fluid
four-velocity $\vct{u}$ and magnetic field $\vct{B}$. The relative
squared norm error for the velocity is computed as
$(\norm{\hat{\vct{u}}}^2 - \norm{\vct{u}}^2)/\norm{\vct{u}}^2$, where
$\hat{\vct{u}}$ is the interpolated and $\vct{u}$ the known reference
value, and $\norm{\vct{u}}=\sqrt{\abs{g_{ab}u^a u^b}}$.  By relative
spatial error we refer to
$(\norm{\hat{\vct{u}}_s-\vct{u}_s}^2)/\norm{\vct{u}_s}^2$, where
$\vct{u}_s$ is vector $\vct{u}$ with a zero time component. The errors
for the magnetic field $\vct{B}$ are computed similarly. For the scalars
density and internal energy, Figure~\ref{fig:grmhd_plots} shows the usual
absolute relative difference with respect to the known values. In the
Figure, we have also plotted histograms, and indicated
the median, mean and standard deviations of all the relative errors.

\begin{figure}
    \begin{center}
        \includegraphics[width=\columnwidth]{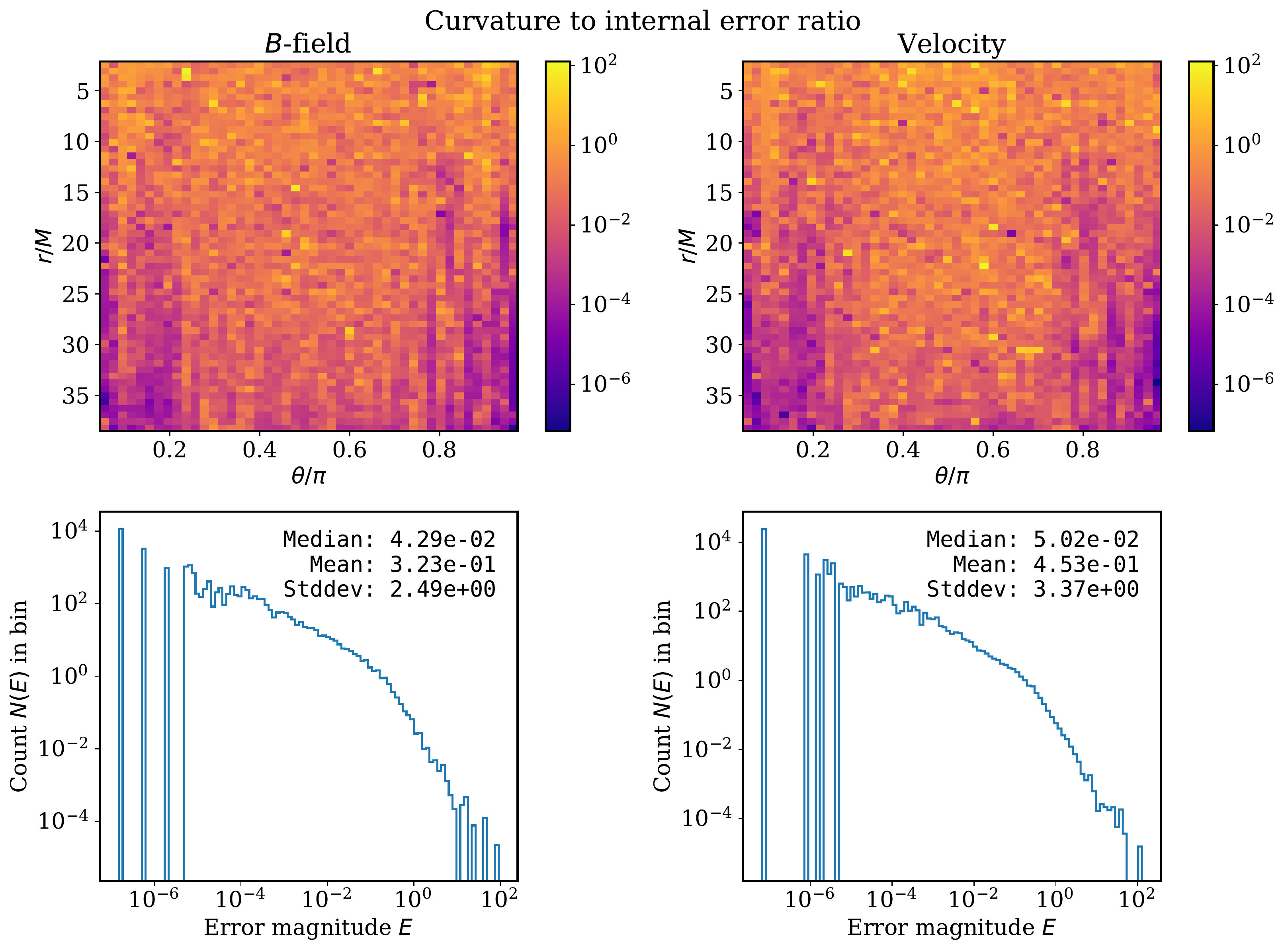}
        \includegraphics[width=0.85\columnwidth]{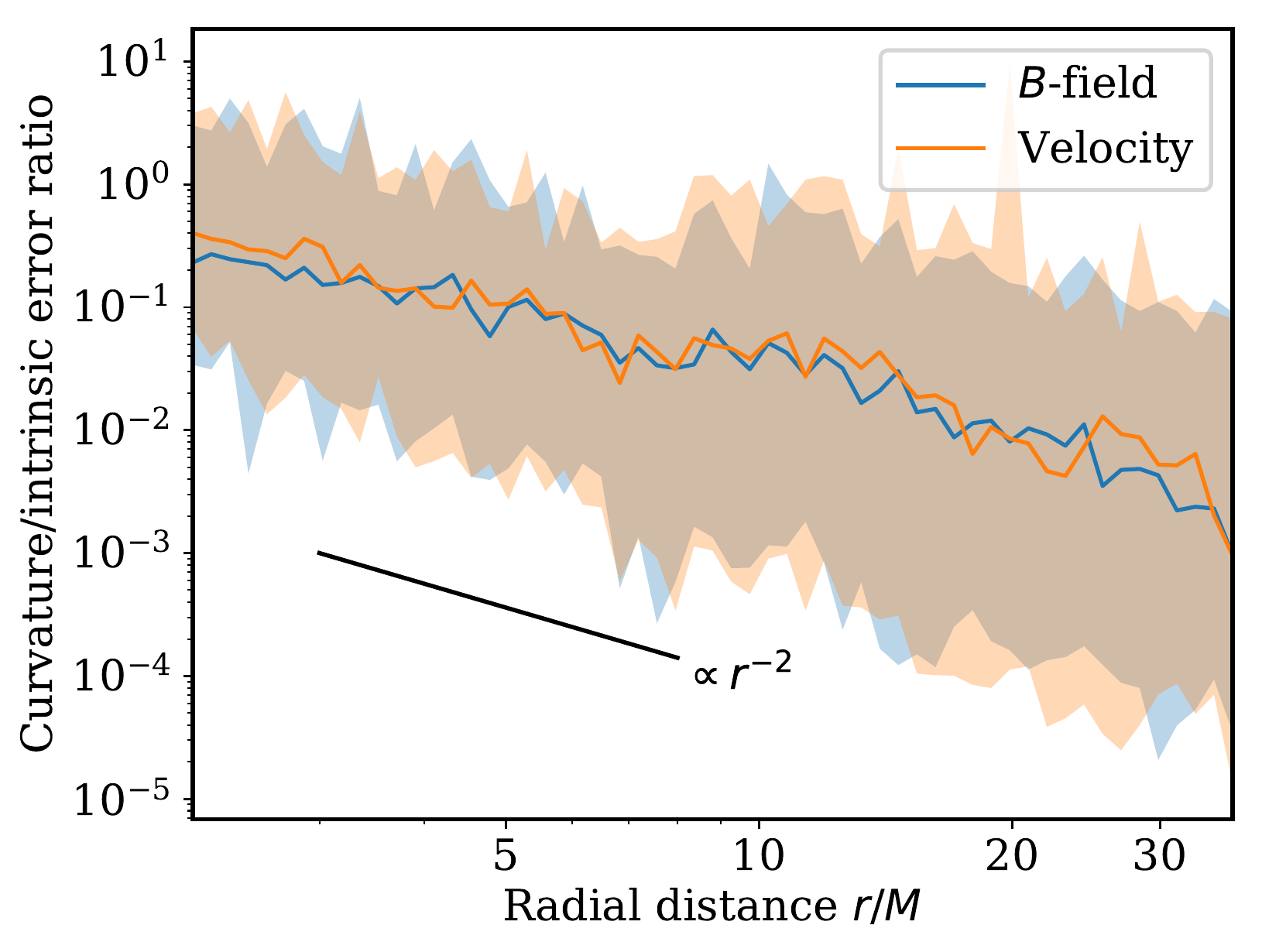}
        \caption{Top panel: Heatmaps (top) and distributions (bottom) of the ratio of the
        curvature error to the intrinsic variability error,
        equation~\eqref{eq:error_ratio}. The leftmost two panels show the result
        for the magnetic field and the rightmost two panels show the
        result for the velocity field.
        Bottom panel: The median ratio of the curvature error to
        the intrinsic variability error as a function of the radial
        distance $r$ from the black hole. Shaded area indicates the limits
        of one Median Absolute Deviation. A black line segment proportional to
        $r^{-2}$ has been added to help guide the eye.
        }
        \label{fig:error_ratios}
    \end{center}
\end{figure}

In Figure~\ref{fig:error_ratios}, we have plotted separately
from equation \eqref{eq:flat-2nd} the ratio $\eratio$ between the squared norm of
the interpolation error due to the curvature terms and the error due
to intrinsic variability, as represented by the covariant second
derivative. For velocity, we computed
\begin{equation}\label{eq:error_ratio}
    \eratio(u) = 
    \frac{\norm{\frac{1}{2}C^a_{bc}(u)\sum_i \chi_i \Delta x^b_i \Delta x^c_i}^2}{%
        \norm{\hat{u}^a - u_0^a - \frac{1}{2}C^a_{bc}(u)\sum_i \chi_i \Delta x^b_i \Delta x^c_i}^2
    },
\end{equation}
where the denominator is equivalent to 
$u^a_{;bc}\sum_i \chi_i \Delta x^b_i \Delta x^c_i$. The ratio $\eratio(B)$
for magnetic field was computed similarly. As in
Figure~\ref{fig:grmhd_plots}, the errors have been averaged over the
azimuthal angle $\phi$.


From Figure~\ref{fig:error_ratios}, we see
that for at least this particular GRMHD simulation, the
strong internal variability of the velocity field and the magnetic field
dwarfs the curvature effects on the interpolation error. When the bulk
of the simulation volume is considered as a whole, the median curvature
contribution to the error is only on the order of $\sim 5\%$ of the
error caused by the strong intrinsic variability. This is not
unexpected, since the magnetized torus rotating around the black hole is
subject to the magneto-rotational instability (MRI) \citep{balbus1991}, and
will develop strong turbulence and corresponding intrinsic variability.
However, even in this case of strong intrinsic variability, we can see
from Figure~\ref{fig:error_ratios} that when regions closer to the black hole
are considered, the curvature errors can surpass intrinsic
variability errors. This is potentially important, since in the light of
the recent Event Horizon Telescope (EHT) results \citep{ehtI}, it is
precisely these
regions near the black hole event horizon that are expected to be
most interesting. This is due to the fact that in the EHT data, the observable
ring-like feature, which is used to constrain the black hole mass and spin,
and attempt to separate General Relativity from other gravitational
theories, is found in the immediate vicinity of the black hole event
horizon, at $\sim 11~M$ \citep{ehtVI}. From
Figure~\ref{fig:error_ratios}, we see that the curvature error at this
distance might be expected to be around $10\%$ of the intrinsic magnetic
or velocity field error, albeit with a large scatter. This indicates
that when interpolating future observational data at these distances,
accounting for the spacetime curvature can potentially be important.

The results in Figure~\ref{fig:error_ratios}
might conceivably have some dependence on the resolution of the GRMHD
simulation. However, from equation~\eqref{eq:flat-2nd} we see that the
possible resolution dependence of the curvature to intrinsic error ratio
can only emerge from the ratios between vector field component
magnitudes, and their first and second covariant
derivatives.
In a simulation such as the one used in this work, the simulation grid can be
assumed to represent an implicit block filter, which is convolved with
the 'true' field to yield an effective large eddy simulation \citep[e.g.][]{miesch2015}.
In this case, \emph{if} the true smallest variability scales are much smaller
than the grid size, the ratios of the maximum values of the vector field
components and their derivatives should be independent of the grid
resolution.
Correspondingly the ratio between the maximum intrinsic and curvature
errors should be resolution independent as well, as long as these
assumptions hold. However, it should be noted that an increase in the
resolution of the simulation will naturally always decrease the
\emph{total} absolute interpolation error.

Finally, as can be expected from the results in Figure~\ref{fig:error_ratios},
the interpolation errors for both vector quantities, velocity and
magnetic field, shown in Figure~\ref{fig:grmhd_plots}, are quite similar
for both methods. For the scalar quantities, density and internal
energy, the results are likewise nearly identical.
On the whole, the curved-space barycentric method gives more accurate
results in the case of high intrinsic variability
as well, but the improvement in the median error
is only on the order of percents. This is a much more modest improvement
when compared to the locally constant field in
Section~\ref{sc:locally_const}.

\section{Conclusions}

We have presented a covariant generalization of the barycentric
interpolation method, suitable for constrained or unconstrained data on
Riemannian and semi-Riemannian manifolds. The method is based on
computing barycentric coordinates of an interpolation point using
Riemann normal coordinates and parallel transport of all data to the
interpolation point before computing the interpolant. The same approach
also allows interpolation of constrained data without violating the
constraint.

We have shown that the new method attains the linear precision property
of barycentric interpolation in a coordinate-invariant sense, whereas
the coordinate-only method is unable to replicate accurately even
locally constant vector or tensor fields. This property was
demonstrated in practice by interpolating data sampled from a locally
constant vector field defined in a Kerr spacetime. The results showed
that for interpolation regions ranging over four orders of magnitude in
edge length $L$, the method presented here gave exact results up to
floating point precision, whereas the coordinate-only method had an
error proportional to at least $L^2$.

We further investigated the performance of the new method in the context
of a General Relativistic Magnetohydrodynamics simulation, where the
interpolated fields are highly non-linear. Here we saw that even though
the intrinsic variability of the data was high, the barycentric method
still provided improvements on the order of a few percent level.

This paper opens up some interesting new avenues for future work.  An
obvious followup is to investigate numerically efficient implementations
of the new method and to compare the accuracy and computational cost to
existing interpolation methods in a larger numerical survey.  In
addition, for certain geometries and choices of coordinates and/or
vertex positions, the formulae may admit solutions in explicit form.
This would provide an immediate computational speedup for that
particular problem. Finally, there is the suggestion that similar
straightforward covariant generalizations might be found for other known
interpolation methods as well.
Such generalizations might prove particularly useful for GRMHD
simulation codes. This is because typically finite-volume simulations use high-order
interpolation schemes to reconstruct grid cell boundary values from cell
averaged values. High-order methods that take coordinate curvature into
account do exist \citep{mignone2014} and are used in GRMHD simulations
\citep[e.g.][]{white2016}. However, the results in this paper suggest that in
strong gravity situations, the spacetime curvature also needs
to be accounted for.

\section*{Acknowledgements}
We thank the anonymous referee for positive and insightful
comments, which were helpful in producing the final version of this
paper.

This research has made use of NASA's Astrophysics Data System
Bibliographic Services.

The authors acknowledge the financial support of the European Research
Council via ERC Consolidator Grant KETJU (no. 818930).  In addition,
P.P. acknowledges the financial support of the Magnus Ehrnrooth
Foundation.



\bibliographystyle{mnras}
\bibliography{refs} 




\bsp	
\label{lastpage}
\end{document}